\documentclass[aps,twocolumn,pre,showpacs,eqsecnum]{revtex4}
\usepackage{graphpap}
\usepackage[dvips]{graphicx}
\usepackage[dvips]{graphics}
\usepackage{color}

\begin{document}

\title{Raman spectroscopy on mechanically exfoliated pristine graphene ribbons}
 \author{B. Terr\'es$^{1,2}$, S. Reichardt$^{1}$, C. Neumann$^{1,2}$,  K. Watanabe$^3$,  T. Taniguchi$^3$, and C. Stampfer$^{1,2}$}
 \affiliation{%
$^1$JARA-FIT and 2nd Institute of Physics, RWTH Aachen University, 52074 Aachen, Germany, EU\\
$^2$Peter Gr\"unberg Institute (PGI-9), Forschungszentrum J\"ulich, Germany, EU\\
$^3$Advanced Materials Laboratory, National Institute for Materials Science, 1-1 Namiki, Tsukuba, 305-0044, Japan
}

\date{ \today}

 \begin{abstract}
We present Raman spectroscopy measurements of non-etched graphene nanoribbons, with widths ranging from 15 to 160~nm, where the D-line intensity is strongly dependent on the polarization direction of the incident light. The extracted edge disorder correlation length is approximately one order of magnitude larger than on previously reported graphene ribbons fabricated by reactive ion etching techniques. This suggests a more regular crystallographic orientation of the non-etched graphene ribbons here presented. We further report on the ribbons width dependence of the line-width and frequency of the long-wavelength optical phonon mode (G-line) and the 2D-line of the studied graphene ribbons.
 \end{abstract}

 \maketitle

\newpage

\section{Introduction}
Graphene nanoribons have been extensively studied in the past years \cite{Han07,Lin07,Wan08,Mol10,Gal10,Ter11}, mainly due to their promise of an electronic band-gap making them interesting for electronic applications. Confinement of electrons in these nanoscaled structures is predicted to form a quasi one-dimensional system \cite{Lin08} with its properties strongly depending on the configuration of the edges \cite{Son06,Yan07}. However, experimental and theoretical studies have revealed graphene nanoribbons to be extremely sensitive to small amounts of disorder, in particular to edge disorder~\cite{Wan11,Kos09}. In fact, the transport characteristics of nanostructured graphene ribbons are mainly dominated by statistical Coulomb blockade effects~\cite{Gal10,Sta09}. Improvements on the fabrication techniques allowing for cleaner edge configurations 
are therefore of great importance and may not only improve the transport properties~\cite{Tom11} but also enable the investigation of the unique vibrational properties of these graphene nanostructures~\cite{Sai10}. Despite theoretical work~\cite{Gil09,Gil10} there are - to our knowledge - only a few optical characterization studies of graphene nanoribbons~\cite{Bis11,Ryu11}. 
Raman spectroscopy of carbon materials, in general, has been identified as a powerful tool for determining the number of graphene layers \cite{Fer07,Mal09}, the local amount of strain and doping \cite{Lee12}, and for studying electron-phonon interactions \cite{{Yan07,Che11,Sta07,Pis07}} and therefore the electronic properties themselves.

In this work, we report on Raman spectroscopy measurements on non-etched graphene ribbons of various widths (from $\approx$15 to 160~nm) resulting from peeling-off a graphene flake on the boundary region of a hexagonal boron nitride (hBN) flake and its underlying SiO$_{\text{2}}$ substrate. We show that the characteristic signatures of single-layer graphene are well preserved and that the configuration of the edges is more regular compared to previously studied graphene ribbons fabricated by state-of-the-art reactive ion etching (RIE) techniques~\cite{Bis11,Ryu11}. Moreover, the analysis of the full width at half maximum (FWHM) of the G- and 2D-line ($\Gamma_{G}$ and $\Gamma_{2D}$) as well as the frequency of the G- and 2D-line ($\omega_{G}$ and $\omega_{2D}$) 
provide strong indications of finite size and/or edge effects~\cite{Gil09,Gil10,Ryu11}.

\section{Fabrication}

\begin{figure}[t]%
\includegraphics*[draft=false,keepaspectratio=true,clip,%
						width=1\linewidth%
						,height=9.0cm]{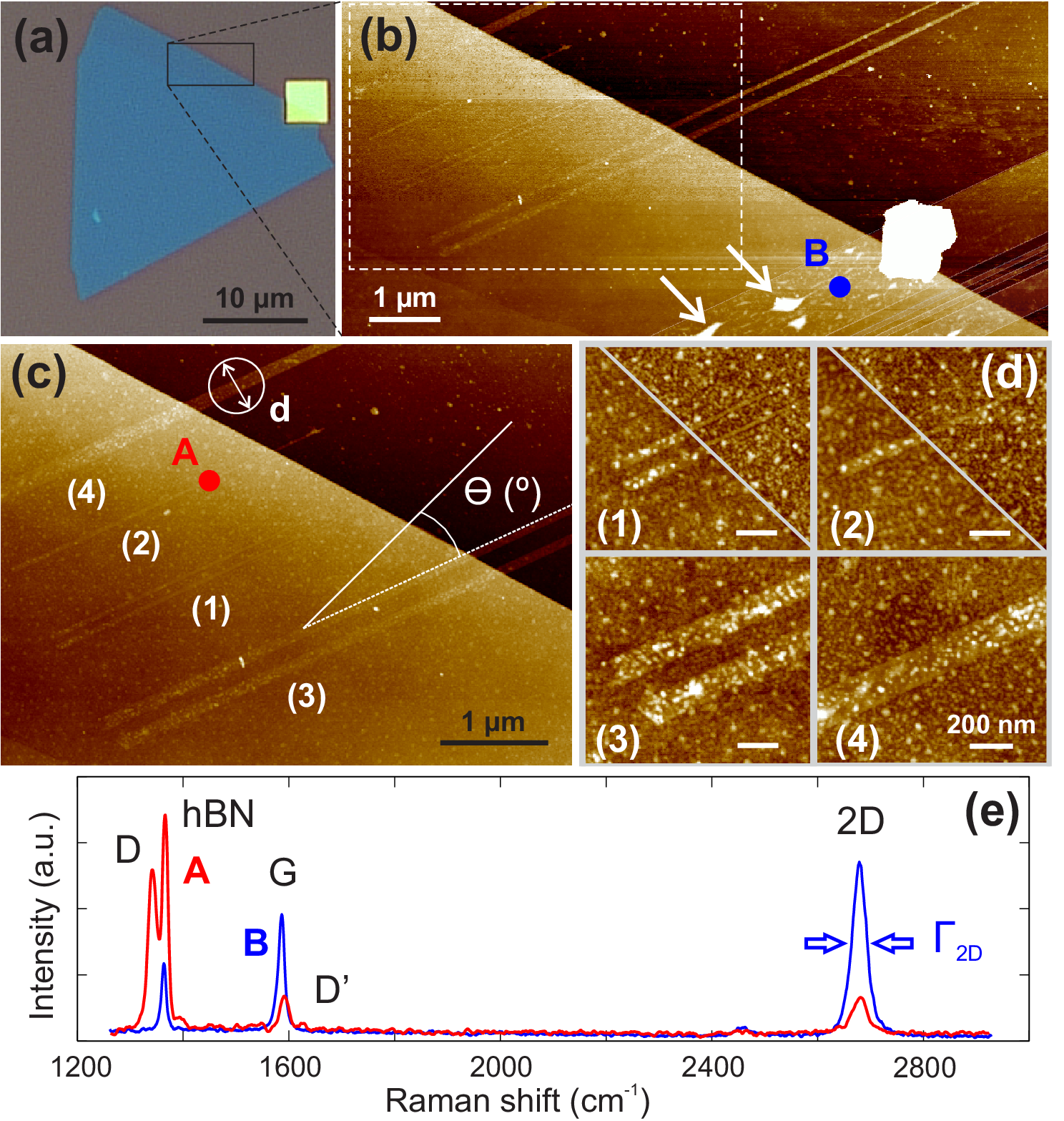}
\caption{%
(color online)
(a) Optical microscope image of a $\approx\!30\,$nm thin hBN flake (light blue color) on top of a Si/SiO$_{\text{2}}$ substrate (grey color). 
(b), (c) and (d) Scanning force microscope (SFM) images taken in the region highlighted by the black box in panel (a). 
(c) SFM close-up image of the white-dashed box in panel (b). In this region the ribbons are separated by a distance of around $1\,\mu$m, twice as large as the spot-size $d\!\approx\!500$ nm (white circle) of the linearly polarized laser with an angle $\theta$. 
(d) SFM close-ups of the ribbons (1), (2), (3) and (4), also displayed in panel (c). 
Ribbons (1) and (2) do not have a constant width, as highlighted in the two upper subpanels of panel (d). We show the wider and narrower ends of these ribbons.
%
(e) Characteristic Raman spectra of bulk graphene on hBN [acquisition point B in panel (b)] and of ribbon (2) [acquisition point A, panel (c)].
}
\label{onecolumnfigure}
\end{figure}

The fabrication of the graphene ribbons is based on purely mechanical exfoliation of graphite. We initially prepared
Si/SiO$_{\text{2}}$ samples with deposited hBN flakes (Fig. 1a). The hBN flakes have been mechanically exfoliated from pure hBN crystals and deposited onto the Si/SiO$_{\text{2}}$ substrate. Thereafter, the samples were immersed in a piranha solution, 3:1 mixture of sulfuric acid (H$_2$SO$_4$) and $30\%$ hydrogen peroxide (H$_2$O$_2$), for 3 minutes and later rinsed with ultrapure water. This cleaning procedure has a similar effect on the SiO$_{\text{2}}$ surface than a plasma etching step prior deposition of the graphene flakes. Both methods are supposed to hydroxylate the SiO$_{\text{2}}$ surface~\cite{Tib13} and therefore increase the local adhesion of graphene to the surface. The Raman spectrum of graphene on such a treated SiO$_{\text{2}}$ substrate is characterized by a very slight increase of the FWHM of the 2D-line~\cite{Wan12}.
The hBN flakes are known to be chemically inert and therefore not affected by the piranha solution at room temperature \cite{Alt07}. Interestingly, we nonetheless observe an increase in doping of graphene on hBN compared to graphene regions resting on SiO$_{\text{2}}$.

\begin{figure}[b]%
\includegraphics*[draft=false,keepaspectratio=true,clip,%
						width=1\linewidth%
						,height=12.0cm]{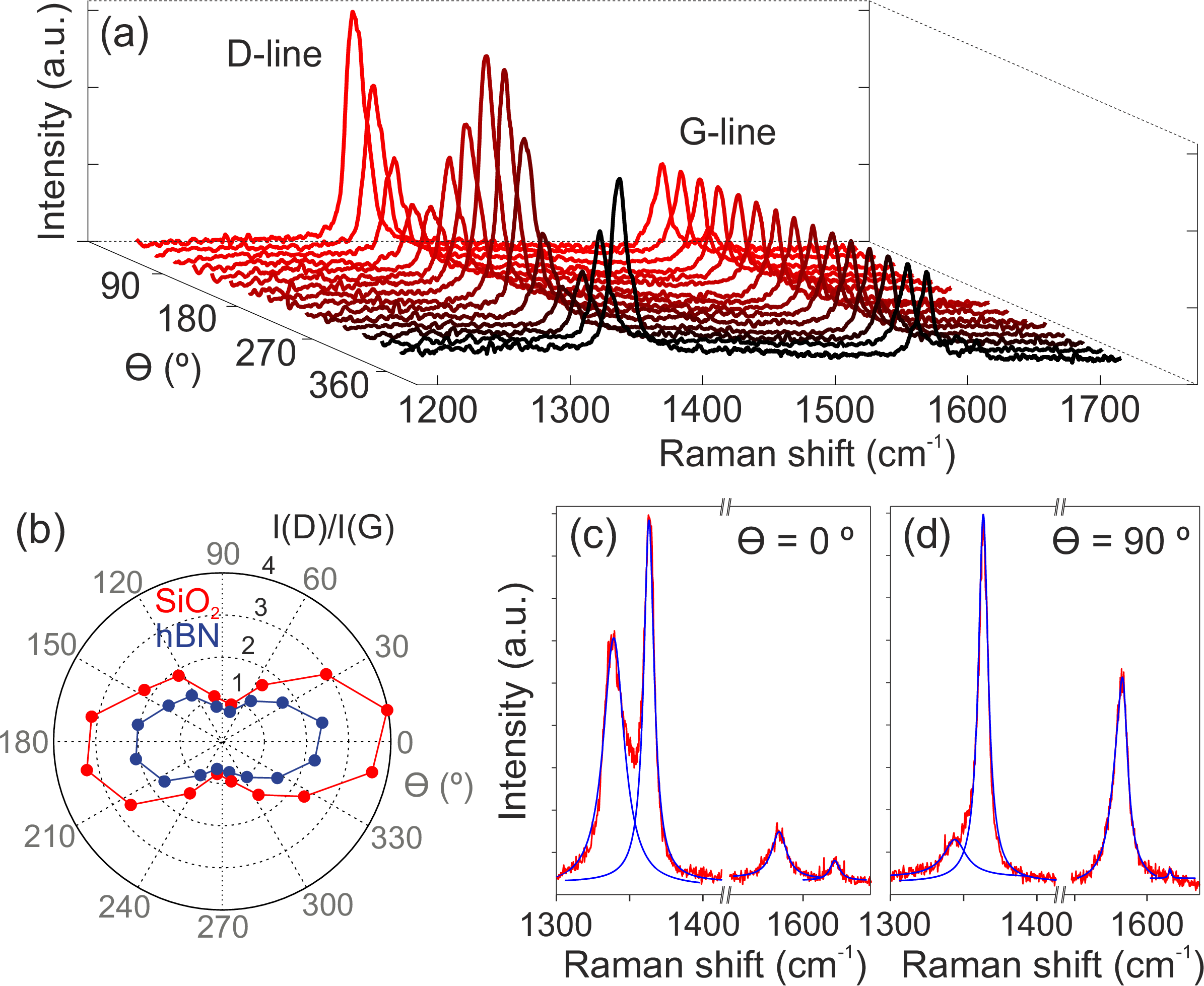}
\caption{%
(color online)
(a) Raman spectra (D- and G-line) of the ribbon (3) on SiO$_{\text{2}}$ as a function of the polarization angle $\theta$ (see Fig. 1c). The difference in polarization angle between subsequent traces is $\theta\!=\!22.5^{\circ}$. The Raman spectra are normalized to the G-line maximum height and shifted for clarity.
(b) Polar plot of $I(D)/I(G)$ as a function of $\theta$ for ribbon (3) on both hBN (blue trace) and SiO$_{\text{2}}$ (red trace) substrates. 
(c) and (d) Raman spectra of ribbon (4), on the hBN substrate, at $\theta\!=\!0^{\circ}$ (c) and $90^{\circ}$ (d). The Lorentzian fits to the data are shown in blue.
}
\label{onecol}
\end{figure}

While the hBN flakes have been directly deposited on the SiO$_{\text{2}}$ substrate, the graphene flakes have been prepared on top of a $\approx$300\,nm-thick layer of polymethylmethacrylate (PMMA) previously spin-coated on a glass slide \cite{Zom11}. Raman spectroscopy was used to identify and select individual single-layer graphene flakes \cite{Fer07,Gra07}. The resulting graphene-PMMA-glass stamp was then mounted in a mask-aligner in such a way that the graphene flake could be aligned on top of the hBN-SiO$_{\text{2}}$ piranha-treated chip \cite{Wan13}. Once on top of the hBN-SiO$_{\text{2}}$ target region, the two flakes were brought into contact.
This process was repeated until some parts of the graphene flake stuck to the hBN-SiO$_{\text{2}}$ surface. 
This technique utilizes van der Waals adhesion to peel-off the graphene ribbons (shown in Fig. 1a), the hBN substrate is therefore important for this fabrication process since graphene adheres more strongly to the hBN than SiO2 \cite{Wan13}. The yield of this fabrication process is nonetheless low and neither the position nor the width of the obtained graphene ribbons is controllable. Therefore, this fabrication method is -~in its present form~- irrelevant from a technological point of view, but it is extremely valuable since it allows the Raman (and potentially transport) investigation of non-etched, i.e. pristine, graphene ribbons. Moreover, we would like to emphasize that these graphene ribbons were never in contact with any spin-coated polymer resist typically involved in the fabrication of etched ribbons, nor with any solvents such as acetone, isopropanol or even water.

An optical microscope image of a fabricated sample is shown in Fig.~1a. For simplicity, we grouped the ribbons of similar width and labeled them as (1)-(4) (shown in Fig.~1c). The widths were extracted from scanning force microscope (SFM) images (Figs.~1b, 1c and 1d), resulting in a width of W $\approx$ 160 and 120 nm for the ribbons (4) and (3). The widths of the ribbons (1) and (2) differ significantly between left and right ribbon ends (see upper panels in Fig.~1d). Specifically, ribbons (1) and (2) have a varying width from W $\approx$ 40 to 15~nm [ribbon (1)] and W $\approx$ 45 to 20~nm [ribbon (2)].
In the following, we will therefore refer to the average width $W\approx$ 25, 35, 120 and 160~nm of the ribbons (1), (2), (3) and (4). The Raman data were recorded using a laser wave length of 532~nm ($\hbar \omega_L\!=\!2.33\,eV$) through a single-mode optical fiber whose spot size is limited by diffraction. The measurement setup is a commercially available confocal Raman Microscope Alpha 300R from Witec, whose laser is linearly polarized. The sample was fixed to a high-precision rotation mount model PRM-1 from Thorlabs, in order to manually adjust the polarization laser direction relative to the ribbon axis (see inset in Fig. 1c). A long working distance focusing lens with numerical aperture of 0.85 is used to obtain a spot size of approximately 500 nm (circle in Fig. 1c). Characteristic Raman spectra of the narrow ribbon (2) and bulk graphene, both on the hBN substrate, are presented in Fig.~1e. The Raman spectra (labels A and B in Fig.~1b and 1c) show the prominent G-line ($\approx$1582 cm$^{-1}$) as well as the single Lorentzian-shaped 2D-line ($\approx$2675 cm$^{-1}$) as expected for graphene. No defect induced D-line ($\approx$1340 cm$^{-1}$) or D'-line ($\approx$1620 cm$^{-1}$) are observed on the bulk graphene region (acquisition point B), which confirms that the fabrication method does not induce defects into the graphene flake. In both spectra, a third prominent sharp peak arises at $\approx$1365 cm$^{-1}$, which is attributed to the Raman-active LO phonon of hBN~\cite{Gei66}.

\begin{figure}[t]%
\includegraphics*[draft=false,keepaspectratio=true,clip,%
						width=1\linewidth%
						,height=9.0cm]{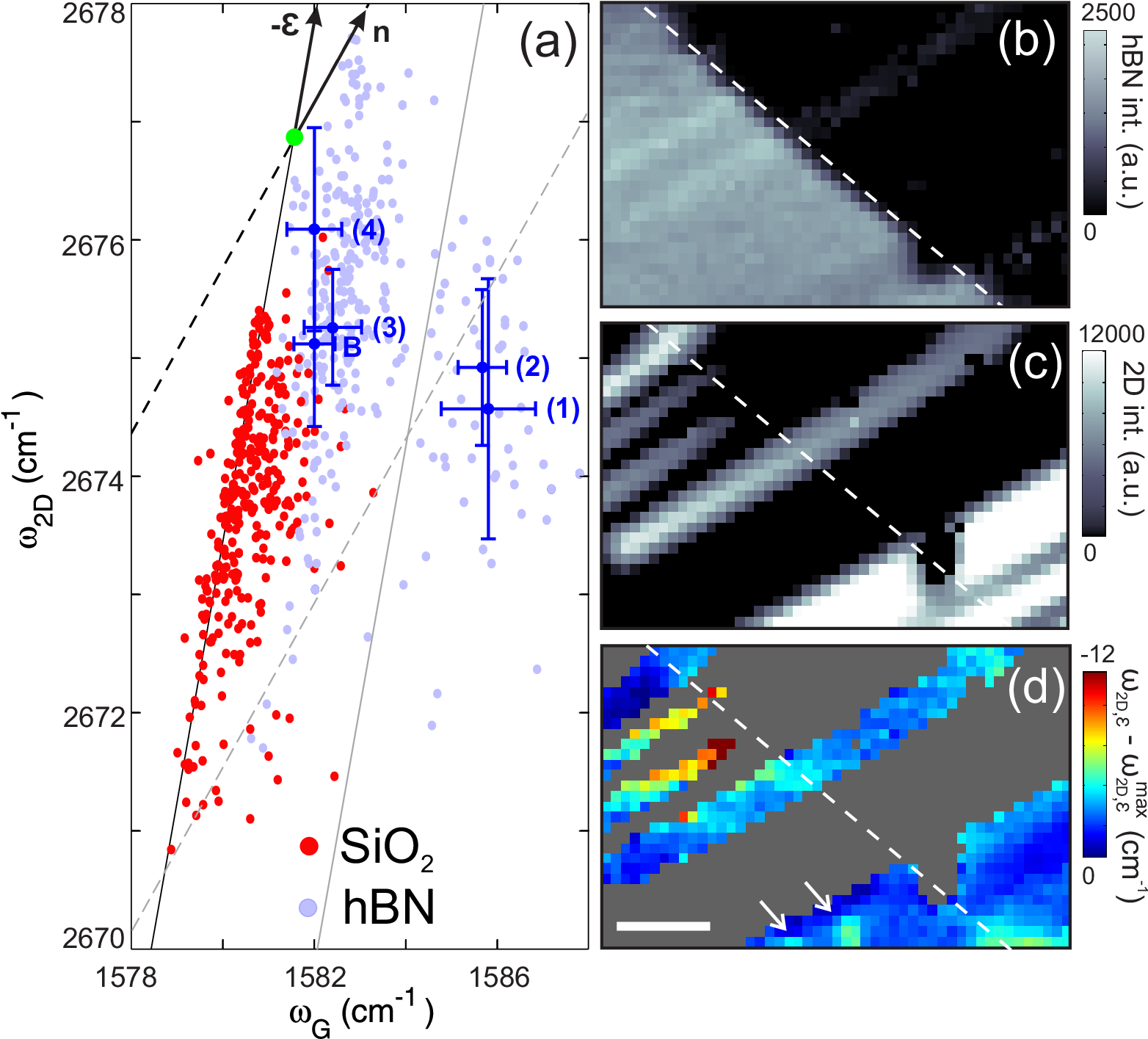}
\caption{%
(color online)
(a) Correlation between $\omega_{2D}$ and $\omega_{G}$ at T = 300 K. The description of the black and gray axis as well as the color code are introduced in the main text.
(b) and (c)  False-colored Raman maps of I(hBN) and I(2D). 
(b) The boundary between the hBN and SiO$_{\text{2}}$ substrates is marked with a white dashed line. 
(c) The individual ribbons (1), (2), (3) and (4) are well differentiated from each other.  
(d) Map of the local 2D-line shifts due to strain $\omega_{2D,\varepsilon}$
 obtained after projecting all the Raman data points onto the strain axis [solid black line in panel (a)] relative to its maximum value [$\omega_{2D,\varepsilon}^{max}$, green point in panel (a)]. The scale bar is $2\mu$m.
}
\label{onecol}
\end{figure}

\section{Characterization of the edges}

In order to characterize the edges and in particular the edge roughness of the graphene ribbons, we performed polarization angle dependent Raman measurements. Fig. 2 shows the Raman spectra of the ribbons (3) (W $\approx$ 120 nm, Figs.~2a and 2b) and (4) (W $\approx$ 160 nm, Fig. 2c and 2d) as function of the polarization angle $\theta$ of the incident light (see inset in Fig.~1c). For each ribbon and each polarization angle, a spectrum has been recorded and the G-, D- and hBN-lines were fitted with a single Lorentzian line shape (see examples in Figs.~2c and 2d). In agreement with previous work~\cite{Bis11,Can04,Gr�03,Cas09}, the D-line intensity $I(D)$ appears to be strongest for polarization parallel to the edge and reaches a minimum for the perpendicular polarization $\theta\!=\!90^{\circ}$. This can be observed in Fig. 2a, where each Raman spectrum corresponds to a different polarization angle $\theta$, starting from $\theta\!=\!11.25^{\circ}$ to $\theta\!=\!348.75^{\circ}$ in steps of $22.5^{\circ}$. Every trace in this plot is normalized to the maximum intensity of the G-line and shifted in the intensity and frequency axis for clarity.
For the rest of the analysis, we compare the ratio $I(D)/I(G)$ using the peak area of the fitted Lorentzian function as a measure of intensity. 
In Fig. 2b we show a corresponding polar plot which illustrates the expected mirror planes at $\theta\!=\!0^{\circ}$ and $\theta\!=\!90^{\circ}$ ~\cite{Can04,Gr�03}.
Raman spectra with Lorentzian fits for the direction of maximum and minimum D-line intensity ($\theta\!=\!0^{\circ}$ and $\theta\!=\!90^{\circ}$, respectively) of ribbon (4) are 
presented in Figs.~2c and 2d. 

According to Ref.~\cite{Cas09} and assuming that $I(G)$ does not depend on $\theta$, a lower bound for the edge disorder correlation length $\xi\!\approx\!2v_F/(\omega_Lb)$ can be estimated from the ratio $b\!=\!I(D)_{min}/I(D)_{max}$ between the lowest and highest normalized D-line intensity ($I(D)_{min}/I(G)$ and $I(D)_{max}/I(G)$, respectively). 
For the ribbon (4) (Fig. 1c and 1d), we obtain the lowest intensity ratio of $b\,\approx\,0.055$ (Fig. 2c and 2d), which yields a correlation length of $\xi\,\approx\,10\,$nm. This value is significantly higher than the correlation length of $\xi\!\approx\!1\,$nm reported on plasma etched graphene nanoribbons~\cite{Bis11} and therefore suggests that the graphene ribbons have a more regular crystallographic orientation of the edges.

\section{Strain, doping and finite size effects}

For a more detailed investigation of the dependence of the Raman spectra
on the width of the graphene ribbons, 
we study
spatially resolved Raman maps of the sample.
%
In particular, we recorded a Raman map of the 6 $\mu$m by 10 $\mu$m sample region shown in Fig. 1b with a spatial oversampling of 200~nm and an integration time of 15~s 
(with a laser spot size of 500~nm and a laser power of $\approx$\,1~mW). The corresponding Raman maps of the hBN-line and the 2D-line intensities, $I(hBN)$ and $I(2D)$, are shown in Figs.~3b and 3c. One can identify the hBN and SiO$_{\text{2}}$ substrates and the graphene ribbons (1)-(4), partly crossing both substrates. In the lower right
corner of Fig.~3c, bulk graphene is also observed.
By means of the so-called vector decomposition method introduced by Lee et al. \cite{Lee12}, 
we analyze the presence of strain and/or doping variations in our sample.
Accordingly, we plot the dependence of the G-line ($\omega_{G}$) and the 2D-line ($ \omega_{2D}$) positions (i.e. frequencies) for all the Raman spectra recorded in the inspected area (Fig. 1b) in Fig.~3a. The red points show the extracted Raman data from spectra recorded on bulk graphene and ribbons, both on SiO$_{\text{2}}$, whereas the light blue points are from graphene regions resting on hBN. The blue data points with error bars show the average values of $\omega_{G}$ and $\omega_{2D}$ obtained for every individual graphene ribbon (1)-(4) and bulk graphene (B) on the hBN substrate (see labels in Fig. 3a). 
\begin{figure}[t]%
\includegraphics*[draft=false,keepaspectratio=true,clip,%
						width=1\linewidth%
						,height=12.0cm]{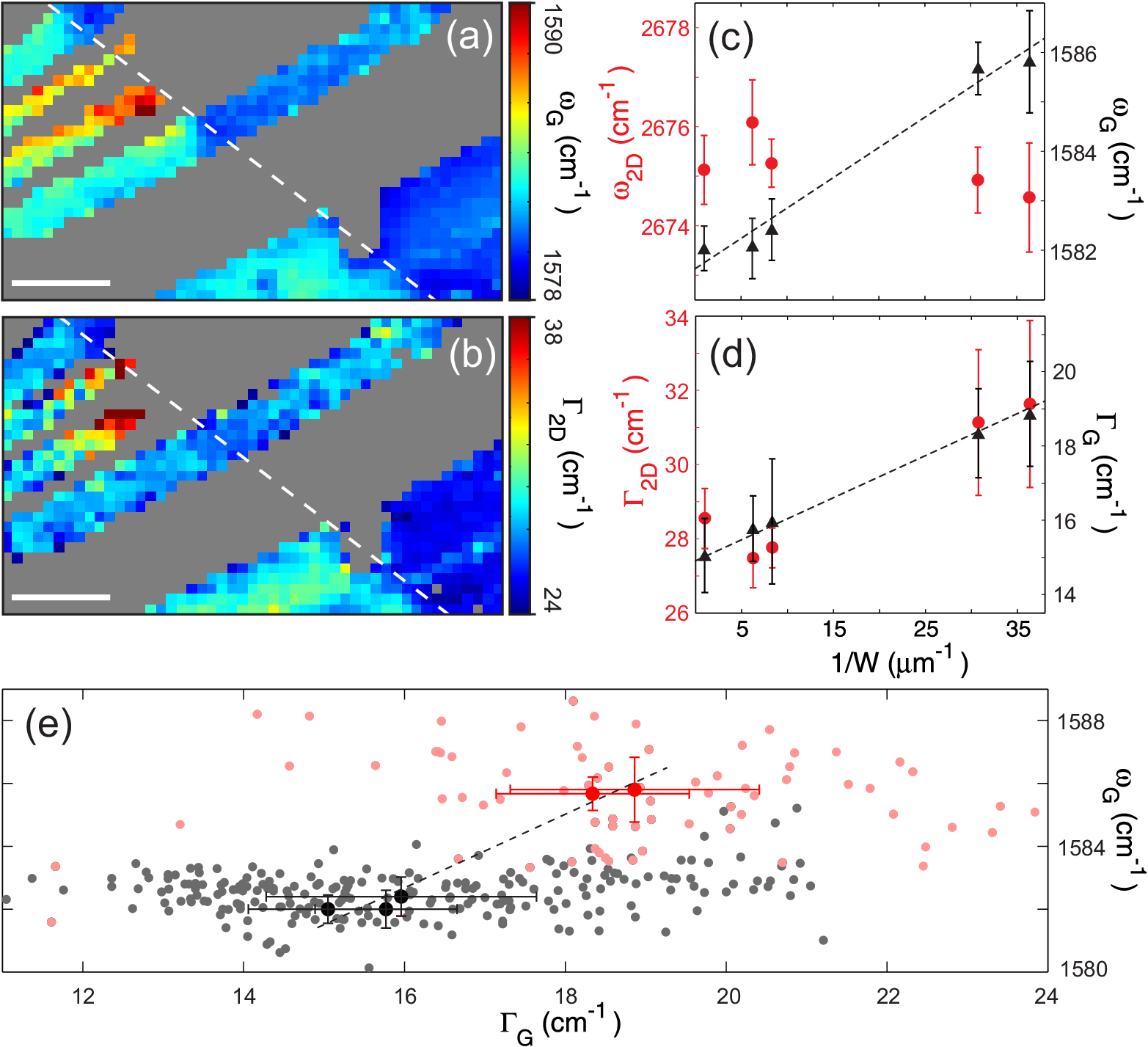}
\caption{%
(color online)
(a) and (b) Local distribution of $\omega_{G}$ and $\Gamma_{2D}$, respectively. The scale bars are 2$\mu$m. 
(c) Averaged $\omega_{G}$ and $\omega_{2D}$ for every individual graphene ribbon on hBN as function of 1/$W$. 
(d) Averaged $\Gamma_{G}$ and $\Gamma_{2D}$ for the individual graphene ribbons on hBN as function of 1/$W$.
(e) Correlations between $\Gamma_{G}$ and $\omega_{G}$ for the ribbons and bulk graphene on the hBN substrate. The light red data points correspond to the narrowest ribbons (1) and (2) with their respective averages marked in red. The ribbons (3), (4) and bulk graphene (B) appear as gray data points with their average values in black. The error bars in all the panels are half the standard deviation.
}
\label{onecol}
\end{figure}
The solid and dashed lines indicate the slopes of the strain and large-scale doping axis according to Ref.~\cite{Lee12}. Please note that we do not know the exact origin of these two axis and, for simplicity, we marked the same origin as in
Ref.~\cite{Lee12} (see green point in Fig. 3a).
Interestingly, the red cloud of data points clearly follows a slope of $\Delta \omega_{2D}/\Delta\,\omega_{G}\!=\!2.2$ (solid black line), characteristic of uniaxial strain - in good agreement with Lee et al. \cite{Lee12} -.
Both red and main light blue data clouds appear to be offset by $\approx\!2.2\,$cm$^{-1}$ in the $\omega_{G}$ axis with a direction parallel to the strain axis. This offset is understood as a difference in induced doping \cite{Che11} between the SiO$_{\text{2}}$ and the hBN substrates (extracted doping difference: $\Delta n\approx 2\,$x$\,10^{11}\,$cm$^{-2}$), most likely due to the treatment with the piranha solution of the hBN surface.
More interestingly, Fig.~3a suggests that the narrowest ribbons [(1) and (2)] are subject to stronger doping compared to bulk graphene and the wider ribbons. This is noticeable from their mean positions [labeled (1) and (2) in Fig.~3a], which are at the very right of this plot (see right gray dashed line of slope 2.2). 
However, there is an inconsistency with the line width of the G-peak that will be discussed in the following section.
Interestingly, the same ribbons [(1) and (2)] seem to have also different strain values compared to bulk graphene and the wider ribbons [(3) and (4)] (see lower dashed line). 
This finding is highlighted after projecting all ($\omega_{2D}$; $\omega_{G}$) points onto the strain axis (the obtained values are labeled as $\omega_{2D,\epsilon}$).
In Fig. 3d we show the corresponding spatial map of the difference $\omega_{2D,\epsilon} - \omega_{2D,\epsilon}^{max}$ relative to its maximum value $\omega_{2D,\epsilon}^{max}$. Here, we show that the strongest deviations are clearly for the two most narrow ribbons (see yellow and red regions in Fig. 3d). Please note that in bulk graphene, the values decrease close to the hBN edge and on bubbles (marked by white arrows in Figs.~1b and 3d), which is a further sign that this quantity is indeed related to strain.



For a more quantitative analysis of the dependence of the Raman G- and 2D-modes on the ribbon width, we analyze the changes in frequency and broadening of the G-line as well as $\omega_{2D}$ and $\Gamma_{2D}$ as a function of the averaged ribbon width $W$. Apart from the aforementioned difference in doping between the hBN and SiO$_{\text{2}}$ substrates (Fig.~3a), the spatial representation of $\omega_{G}$ (Fig.~4a) reveals a stiffening of the G-line for the narrower ribbons (1) and (2), which is in agreement with Fig. 3a and earlier work~\cite{Ryu11}. 
Fig.~4c shows $\omega_{G}$ and $\omega_{2D}$ as a function of the inverse averaged width ($1/W$) for the different ribbons. 
Interestingly, we observe an increase of $\omega_{G}$ as function of $1/W$ (see dashed line in Fig.~4c), meaning that the smaller the ribbon the stiffer the G-line. This is commonly attributed to edge doping and/or confinement effects~\cite{Ryu11}. The 2D-line frequency $\omega_{2D}$ does not show any substantial dependence with the width of the ribbons (see red data points in Fig.~4c). In Fig. 4d we show that also the G- and 2D-peak line widths ($\Gamma_{G}$ and $\Gamma_{G}$) increase with decreasing ribbon width $W$. This width dependent broadening might be an indication
of finite size effects~\cite{Fer00}. 
In order to exclude doping effects for the increase of $\Gamma_{G}$, we show the dependence of $\omega_{G}$ as function of $\Gamma_{G}$ (re-plotting the data shown in Figs.~4c and 4d), in Fig.~4e. 
In complete disagreement with experimental results on bulk graphene~\cite{Yan07,Sta07,Pis07} and theory~\cite{And06} on doping dependent Landau
damping, we observe an increase of $\omega_{G}$ with increasing $\Gamma_{G}$. For bulk graphene, exactly the opposite 
has been observed in earlier experiments~\cite{Yan07,Sta07,Pis07}.
Finally, from Figs.~3a and 3d we can estimate a maximum strain difference in the narrow ribbons. Assuming uniaxial strain, we extract a maximum strain difference on the order of 0.23\%~\cite{Lee12}. It is important to note that according to Ref.~\cite{Moh09} these values cannot explain the observed maximum broadening of the G-line (Fig. 4e), making edge effects and/or finite size effects a prime candidate to explain our experimental findings.    



\section{Conclusion}

In summary, we  discussed Raman spectroscopy measurements on lithography-free fabricated graphene ribbons made by direct exfoliation of graphene on hBN flakes. Despite a prominent doping of the hBN substrate, most probably induced by the fabrication process, we were able to perform polarization dependent measurements that confirm a more regular crystallographic orientation of the ribbon edges. Analysis of the frequency and broadening of the G- and 2D-line show prominent differences between the narrowest ribbons ($\approx$ 15 and 20 nm) and the widest ones (bulk graphene included), suggesting the presence of confinement and/or edge effects in these narrow structures. The results of this work highlight that further developments in the fabrication process yielding cleaner graphene samples with regularly oriented edges may enhance both the vibrational and electronic characteristics of graphene devices.

{Acknowledgment ---}
We thank G. Rossell�, D. May and P. Nemes-Incze for valuable discussions.
Support by the HNF, JARA Seed Fund, the DFG (SPP-1459 and FOR-912), the ERC (GA-Nr. 280140) and the EU project Graphene Flagship (contract no. NECT-ICT-604391) are gratefully acknowledged.


\begin{thebibliography}{99}


\bibitem{Han07}
M.\,Y.\,Han et al.,
Phys. Rev. Lett. \textbf{98}, 206805 (2007).

\bibitem{Lin07}%
Y.\,-M.\,Lin et al.,
Phys. Rev. B \textbf{78}, 161409(R) (2008).
 
\bibitem{Wan08}%
X.\,Wang et al.,
Phys. Rev. Lett. \textbf{100}, 206803 (2008).
 
\bibitem{Mol10}%
F.\,Molitor et al.,
Phys. Rev. B \textbf{79}, 075426 (2010).
 
\bibitem{Gal10}%
P.\,Gallagher et al.,
Phys. Rev. B \textbf{81}, 115409 (2010).

\bibitem{Ter11}%
B.\,Terr\'es, et al., 
Appl. Phys. Lett. \textbf{98}, 032109 (2011). 

\bibitem{Lin08}
Y.\,Lin, et al., 
Phys. Rev. B \textbf{78}, 161409(R) (2008).

\bibitem{Son06}%
Y.\,-W.\,Son, et al., 
Phys. Rev. Lett. \textbf{97}, 216803 (2006).

\bibitem{Yan07}%
J.\,Yan, et al., 
Phys. Rev. Lett. \textbf{98}, 166802 (2007).

\bibitem{Wan11}%
X.\,Wang, et al., 
Nature Nanotechnology \textbf{6}, 563�567 (2011)
 
\bibitem{Kos09}%
D.\,V.\,Kosynkin, et al., 
Nature \textbf{458}, 872-876 (2009).
 
\bibitem{Sta09}%
C.\,Stampfer, et al., 
Phys. Rev. Lett. \textbf{102}, 056403 (2009).
 
\bibitem{Tom11}%
N.\,Tombros, et al.,
Nature Physics\textbf{7}, 697-700 (2011).

\bibitem{Sai10}%
R.\,Saito, et al., 
J. Phys.: Condens. Matter \textbf{22}, 334203 (2010).
 
\bibitem{Gil09}%
R.\,Gillen, et al., 
Phys. Rev. B \textbf{80}, 155418 (2009).
 
\bibitem{Gil10}%
R.\,Gillen, et al., 
Phys. Rev. B \textbf{81}, 205426 (2010).

\bibitem{Bis11}%
D.\,Bischoff et al.,
J. Appl. Phys. \textbf{109}, 073710 (2011).

\bibitem{Ryu11}%
S.\,Ryu et al.,
ACS Nano \textbf{5}, 4123-4130 (2011).

\bibitem{Fer07}
A.\,C.\,Ferrari, 
Solid State Commun. \textbf{143}, 47 (2007).

\bibitem{Mal09}
L.\,M.\,Malard, et al., 
Phys. Rep. \textbf{473}, 51 (2009).

\bibitem{Lee12}%
J.\,E.\,Lee, et al., 
Nature Commun. \textbf{3}, 1024 (2012).

\bibitem{Che11}
 C.\,-F.\,Chen, et al., 
Nature \textbf{471}, 617-620 (2011).




\bibitem{Sta07}%
C.\,Stampfer, et al., 
Appl. Phys. Lett. \textbf{91}, 241907 (2007).





\bibitem{Pis07}%
S.\,Pisana, et al., 
Nature Materials \textbf{6}, 198-201 (2007).

\bibitem{Tib13}%
A.\,Tiberj et al.,
Scientific Reports \textbf{3}, 2355 (2013).

\bibitem{Wan12}%
Q.\,H.\,Wang et al.,
Nature Chemistry \textbf{4}, 724-732 (2012).
 
\bibitem{Alt07}%
A.\,O.\,Altun et al.,
Nanotechnology \textbf{18}, 465302 (2007).

\bibitem{Zom11}%
P.\,J.\,Zomer et al.,
Appl. Phys. Lett. \textbf{99}, 232104 (2011).

\bibitem{Gra07}
D.\,Graf et al.,
Nano Lett. \textbf{7}, 238 (2007).
 
\bibitem{Wan13}%
L.\,Wang et al.,
Science \textbf{342}, 614-617 (2013).
 
\bibitem{Gei66}%
R.\,Geick et al.,
Phys. Rev. \textbf{146}, 543 (1966).
 
\bibitem{Can04}%
L.\,G.\,Can�ado et al.,
Phys. Rev. Lett. \textbf{93}, 247401 (2004).
 
 
\bibitem{Gr�03}%
A.\,Gr\"uneis et al.,
Phys. Rev. B \textbf{67}, 165402 (2003).
 
\bibitem{Cas09}%
C.\,Casiraghi et al.,
Nan. Lett. \textbf{9}, 1433-1441 (2009).

\bibitem{Fer00}%
A.\,Ferrari et al.,
Phys. Rev. B \textbf{61}, 14095-14107 (2000).

\bibitem{And06}%
T.\,Ando et al.,
J. Phys. Soc. Jpn. \textbf{75}, 124701 (2006).

\bibitem{Moh09}%
T.\,M.\,G.\,Mohiuddin et al.,
Phys. Rev. B \textbf{79}, 205433 (2009).





 
 
 




\end{thebibliography}
\end{document}